\documentclass[a4paper,11pt]{article}
\pdfoutput=1 

\usepackage{jcappub} 

\usepackage[T1]{fontenc} 

\title{More about a successful vector-tensor theory of gravitation}

\author[a,1]{R. Dale,\note{Corresponding author.}}
\author[b,c]{D. S\'aez}

\affiliation[a]{Departamento de Estad\'{\i}sica, Matem\'atica 
e Inform\'atica, Universidad Miguel Hernandez, \\ Elche, Alicante, Spain}

\affiliation[b]{Departamento de Astronom\'{\i}a y Astrof\'{\i}sica,
Universidad de Valencia,\\ Burjassot, Valencia, Spain}

\affiliation[c]{Observatorio Astron\'{\o}mico,
Universidad de Valencia,\\  E-46980 Paterna, Valencia, Spain}

\emailAdd{rdale@umh.es}
\emailAdd{diego.saez@uv.es}

\abstract{The vector-tensor (VT) theory of gravitation revisited in this article
was studied in previous papers, where 
it was proved that VT works and deserves attention. 
New observational data and numerical codes
have motivated further development which is presented here. 
New research has been planed 
with the essential aim of proving that current cosmological observations, 
including Planck data,
baryon acoustic oscillations (BAO), and so on, may be explained with VT, a theory
which accounts for a kind of dark energy which has the same equation of state as
vacuum. 
New versions of the codes CAMB and COSMOMC have been designed for applications 
to VT, and the resulting versions have been used to get the cosmological parameters 
of the VT model at suitable confidence levels. The parameters to be estimated 
are the same as in
general relativity (GR), plus a new parameter $D$. 
For $D=0$, VT linear cosmological perturbations reduces to those of GR,
but the VT background may explain dark energy. The fits between 
observations and VT predictions lead to non vanishing $|D|$ upper limits at the 
$1\sigma $ confidence level. The value $D=0$ is admissible at this level, but this
value is not that of the best fit in any case.
Results strongly suggest that VT may explain 
current observations, at least, as well as GR; with the advantage that, as it is 
proved in this paper, VT has an additional 
parameter which facilitates adjustments to current observational data.}


\begin{document}
\maketitle
\flushbottom

\section{Introduction}
\label{sec:1}

In this paper, we are concerned with a vector-tensor theory of gravitation (hereafter VT).
It involves two fields: the metric $g^{\mu \nu} $
and the vector field $A^{\mu} $. This theory was studied in previous papers
\citep{dal09,dal12,dal14,dal15}, in which it was proved that
(i) there are no quantum ghosts and classical instabilities, (ii) 
the parametrized post-Newtonian limit is 
identical to that of general relativity (GR), (iii) 
the radius of the black hole horizon deviates with respect to 
that of GR, and the relative deviations may reach values close to 30 per cent,
(iv) the energy density of the VT vector field plays the role of
a cosmological constant, and (v) by using a minimal model -involving seven parameters- 
for the scalar perturbations of the cosmological background, the  
seven years WMAP (Wilkinson Map Anisotropy Probe)
observations and accurate data about Ia supernova luminosities
may be simultaneously explained. 
All this strongly suggests that VT deserves attention. This theory 
must be tested taking into account current observational data.
Where appropriate, the reference \cite{dal14} will be called paper I.

The field equations and the conservation laws of VT, as well as the basic equations 
describing the background universe and its perturbations
were derived in \citep{dal09,dal12,dal14}. Here, the VT foundations and equations 
are briefly summarized by using the following notation criteria:
our signature is (--,+,+,+), Latin (Greek) 
indexes run from 1 to 3 (0 to 3), symbol $\nabla $ ($\partial $) stands
for a covariant (partial) derivative, the antisymmetric tensor $F_{\mu \nu} $
is defined by the relation 
$F_{\mu \nu} = \partial_{\mu} A_{\nu } - \partial_{\nu} A_{\mu }$, 
quantities $R_{\mu \nu}$, $R$, and $g$ are the covariant components of the Ricci 
tensor, the scalar curvature and
the determinant  of the matrix $g_{\mu \nu}$ formed by the covariant components 
of the metric, respectively. 
Units are chosen in such a way that the gravitational constant, $G$, and the speed of light, $c$,
take on the values $c= G = 1$; namely, we use 
geometrized units. 

The VT action is \citep{wil93,wil06}:
\begin{eqnarray}
I &=& \int (  R/16\pi + \omega A_\mu  A^\mu  R
+ \zeta R_{\mu \nu }
A^\mu  A^\nu  -  \nonumber \\
& &
\varepsilon F_{\mu \nu } F^{\mu \nu }  
+\gamma \,\nabla_\nu  A_\mu  \nabla^\nu  A^\mu
+ L_{m} ) \,\sqrt { - g} \,d^4 x \ ,
\label{1.1}
\end{eqnarray} 
where $\omega$, $\zeta $,      
$\gamma$, and $\varepsilon$ are dimensionless parameters.  
The values of $\zeta $      
and $\gamma$ satisfy the condition $\zeta = \gamma $, parameter $\omega$ vanishes, and the pair
($\varepsilon$, $\gamma$) satisfies the inequality $2\varepsilon - \gamma > 0$, which
guarantees the absence of quantum ghosts and unstable modes in VT (see paper I and references cited therein). 
The inequality $\gamma >0$ must be also satisfied to have a positive $A^{\mu} $ energy density 
-in the background universe- which will play the role of vacuum energy 
(see below). From these considerations, it follows 
that the inequalities 
$\varepsilon > \frac {\gamma}{2}>0 $ must be satisfied. 
Moreover, as it was proved in \cite{dal09},  
the parametrized post-Newtonian limits of VT and general relativity (GR)
are identical whatever the values of the pair ($\varepsilon$, $\gamma$) may be.
Tensor $ F_{\mu \nu } $ has nothing to 
do with the electromagnetic field \citep{bm093,dal09,dal14}. 

The VT field equations derived from  
action (\ref{1.1}) may be written as follows:
\begin{equation}  
G^{\mu \nu} = 8\pi G (T^{\mu \nu}_{GR} + T^{\mu \nu}_{VT}) \ ,
\label{fieles_vt}
\end{equation} 
\begin{equation}
2(2\varepsilon - \gamma)\nabla^{\nu} F_{\mu \nu} = J^{^{A}}_{\mu} \ ,
\label{1.3_vt}
\end{equation} 
where $J^{^{A}}_{\mu} = -2 \gamma \nabla_{\mu} (\nabla \cdot A)$ 
with $\nabla \cdot A = \nabla_{\mu} A^{\mu} $, $G^{\mu \nu}$ is the 
Einstein tensor, $T^{\mu \nu}_{GR}$ is the GR energy momentum tensor, and 
\begin{eqnarray}
T^{\mu \nu}_{VT} &=& 2(2\varepsilon - \gamma) [F^{\mu}_{\,\,\,\, \alpha}F^{\nu \alpha}
- \frac {1}{4} g^{\mu \nu} F_{\alpha \beta} F^{\alpha \beta}] \nonumber \\ 
& &
-2\gamma [ \{A^{\alpha}\nabla_{\alpha} (\nabla \cdot A) + \frac {1}{2}(\nabla \cdot A)^{2}\}
g^{\mu \nu} \nonumber \\
& &
-A^{\mu}\nabla^{\nu} (\nabla \cdot A) - A^{\nu}\nabla^{\mu} (\nabla \cdot A)
] \ .
\label{emtee_vt}
\end{eqnarray}  
Equation (\ref{1.3_vt}) leads to the following conservation law 
\begin{equation}
\nabla^{\mu} J^{^{A}}_{\mu} = 0  
\label{confic}
\end{equation}
for the fictitious current $J^{^{A}}_{\mu}$. Moreover, the 
conservation laws $\nabla_{\mu} T^{\mu \nu}_{GR} = 0 $ and  
$\nabla_{\mu} T^{\mu \nu}_{VT} = 0 $ are satisfied by any solution 
of (\ref{fieles_vt}) and (\ref{1.3_vt}).
 
\section{Differential cosmological equations and initial conditions for integration}
\label{sec:2}   
 
 The basic equations describing a flat homogeneous and isotropic background universe 
were derived \citep{dal09,dal12,dal14} by using the 
basic VT equations of section \ref{sec:1}, the Robertson-Walker line element
\begin{equation}
dS^{2} = -dt^{2} + a^{2} (dr^{2} + r^{2} d\theta^{2} + r^{2} \sin^{2}{\theta} d\phi^{2}) \ ,
\end{equation}
and a vector field with covariant components $[A_{0B}(\tau ),0,0,0]$.   
Here and hereafter, $a$ is the scale factor, 
whose present arbitrary value is assumed to be $a_{0} = 1 $,
symbol $t$ ($\tau $) stands for the coordinate (conformal) time,
and the subscript $B$ stands for background. 
Whatever the function
$f $ may be, $f^{\prime} $ and $\dot{f}$ stand for the partial derivative with respect to the 
radial coordinate $r$ and the conformal time $\tau $, respectively.   
Quantities $\rho $ and $p$ are the total density and pressure of
the cosmological fluid. The subscripts $b$, $c$, $\nu $, and $\gamma $ makes 
reference to the baryon, cold dark matter, massless neutrinos, and photons,
respectively, e.g., $\rho_{\gamma}$ is the CMB (cosmic microwave background) energy density.
Finally, function $\rho^{A} $ ($p^{A}$) is the 
contribution to 
the energy density (pressure) corresponding to the vector field $A^{\mu}$.

As it was proved in papers \citep{dal09,dal12,dal14},  
the following equations are satisfied: 
\begin{equation}
3 \frac {\dot{a}^{2}}{a^{2}}= 8 \pi G a^{2} (\rho_{B}+\rho^{A}_{B})
\label{baseq1}
\end{equation} 
\begin{equation}
-2 \frac {\ddot{a}}{a} + \frac {\dot{a}^{2}}{a^{2}} = 8 \pi G a^{2} (p_{B}+p^{A}_{B})  \ ,
\label{baseq2}
\end{equation} 
\begin{equation}
\rho^{A}_{B} = -p^{A}_{B} = \gamma \Xi_{B}^{2} \ ,
\label{eqest_vt}
\end{equation}  
where $\Xi_{B} = (\nabla \cdot A)_{B}$, and
\begin{equation}  
\Xi_{B} = constant = - \frac {1}{a^{2}} [ \dot{A}_{0B} 
+ 2 \frac {\dot{a}}{a} A_{0B}]    \ .
\label{coscons}
\end{equation} 
These equations allow us to find functions $a(\tau)$, $A_{0B}(\tau)$ and $\rho_{B}(\tau)$, 
by using an appropriate equation of state for the cosmological fluid and 
suitable initial conditions (see below).

According to eqs.~(\ref{eqest_vt}) and~(\ref{coscons}), for $\gamma > 0 $,  
the positive energy density $\rho^{A}_{B}$ has the same properties as the vacuum energy $\rho_{V} $. By this reason,
we hereafter write $\rho^{A}_{B} \equiv \rho_{V}$. Moreover, eq.~(\ref{eqest_vt}) allows us to
write the constant $\Xi_{B} $ in terms of other constants; so, one finds
\begin{equation}  
\Xi_{B} = S_{gn} \Big(\frac {\rho_{V}} {\gamma} \Big)^{1/2} \ ,
\label{xib}
\end{equation} 
where $S_{gn} $ only can take on the values $+1$ or $-1$. 

For a given value of $\Xi_{B} $ and suitable initial conditions for $A_{0B}$, 
eq.~(\ref{coscons}) may be numerically solved together with eqs.~(\ref{baseq1}) and~(\ref{baseq2}).
All the initial conditions are taken, in the radiation dominated era, at $z_{in} = 10^{8}$. At 
this redshift, one easily obtains (see paper I) the following initial values of $\tau$
and $A_{0B} $
\begin{equation}
\tau_{in} = \Big( \frac {\dot{a}}{a} \Big)_{in}^{-1}\ , \,\,\,\,\,\, 
(A_{0B})_{in} = - \frac {\Xi_{B}}{5(1+z_{in})^{2}}  
\Big/ \Big( \frac {\dot{a}}{a} \Big)_{in}   
\ .
\label{inib0}
\end{equation}
The ratio $(\dot{a}/a)_{in}$ may be calculated from eq.~(\ref{baseq1}) as in the standard 
model with vacuum energy $\rho_{V} $; hence, this ratio is that obtained 
by the code CAMB \citep{lew00} for standard GR cosmology, which depends on the number of relativistic species
contributing to $\rho_{Bin}$.

CAMB equations for the standard background with cosmological
constant are valid in VT; however, in this last theory 
parameters $S_{gn} $ and $\gamma $ and the new background equation 
(\ref{coscons}) must be included.

Let us now consider tensor, vector and scalar perturbations of the VT background universe,
which were studied in \cite{dal12,dal14} by using
the Bardeen formalism, in which $Q^{(0)}$, $Q^{(1)\pm}_{i}$, and 
$Q^{(2)}_{ij}$ are harmonics (see \cite{bar80,huw97,dal12}) that may be used to expand the scalar, 
vector and tensor perturbations, respectively. 

The covariant components $A_{\mu} $ may be expanded in vector [superscript (1)] 
and scalar [superscript (0)] harmonics as follows:
\begin{equation}
A_{\mu} = (A_{0B}+\alpha^{(0)}Q^{(0)}, \beta^{(0)} Q^{(0)}_{i} + \alpha^{(1)\pm} Q^{(1)\pm}_{i})  \ .
\label{a0b0}
\end{equation}

Evidently,
there are no tensor modes in the $A^{\mu} $ expansion. Therefore, in GR and VT there are the 
same tensor cosmological perturbations (primordial gravitational waves) evolving in the same 
way. The fundamental equation for these perturbations is  
\begin{equation}
\ddot{H}_{_{T}}^{(2)} + 2 \frac {\dot{a}}{a} \dot{H}_{_{T}}^{(2)}
+k^{2} H_{_{T}}^{(2)} = p_{B} a^{2} \Pi_{_{T}}^{(2)} \ ,
\label{gw}                                      
\end{equation}   
where $\Pi_{_{T}}^{(2)} Q^{(2)}_{ij}$ and $H_{_{T}}^{(2)} Q^{(2)}_{ij}$ 
(see \cite{bar80}) are the tensor parts of the anisotropic
stress tensor and the metric, respectively.

The vector modes ${A}^{(1)\pm}$ satisfy the harmonic oscillator equation
\begin{equation}
\ddot{A}^{(1)\pm} + k^{2} {A}^{(1)\pm} = 0      \ ,
\label{a1evol}   
\end{equation}   
whose solutions are well known. Evidently,
these vector modes are not coupled 
to the remaining modes of VT, which coincide with those of GR 
and evolve in the same way due to the fact that,
according to eq.~(\ref{emtee_vt}), the vector part of $T^{\mu \nu}_{VT}$ vanishes.

All the scalar modes of GR are also involved in VT, but new
scalar modes characteristic of VT must be included.
The modes $\alpha^{(0)}$ and $\beta^{(0)}$ of eq.~(\ref{a0b0}) are not 
appropriate. As it was proved in \cite{dal12,dal14}, 
the most suitable VT scalar mode may be defined as follows: 
\begin{equation}
\Xi \equiv \nabla \cdot A = \Xi_{B} (1+ \Xi Q^{(0)}) \ ,
\label{diva_exp}
\end{equation}
which means that the new VT mode is the first order term in the harmonic expansion of the 
scalar function $\nabla \cdot A$.
There are no more independent VT scalar modes associated to field $A^{\mu} $.

Calculations are performed in the synchronous gauge; 
in which, the scalar perturbations corresponding to the metric, the four-velocity, and the
energy-momentum tensor of a cosmological fluid are expanded as follows \citep{mb95}:
\begin{eqnarray}
& &
g_{00}=-a^{2}, \,\,\,\,\,\,\,\, g_{0i} = 0, \,\,\,\,  
\nonumber \\
& & 
g_{ij}=a^{2}[(1+\frac{h}{3}Q^{(0)})\delta_{ij}-(h+6\eta)Q_{ij}^{(0)}]                     
\nonumber \\
& &
U_{i} = \frac{a}{k} \theta Q^{(0)}_{i}, \,\,\,\, \rho=\rho_{B}(1+\delta Q^{(0)})
\nonumber \\
& &
T_{ij} = p_{B}(1+\pi_{L} Q^{(0)})\delta_{ij} + \frac{3}{2}(\rho_{B}+p_{B})\sigma Q_{ij}^{(0)}  \ ,
\end{eqnarray}     
where function $Q^{(0)} = \exp ({i\vec{k} \cdot \vec{r}})$ is a plane wave, 
$Q^{(0)}_{i} = (-1/k) \partial_{i} Q^{(0)}$, and  $Q_{ij}^{(0)} = k^{-2} \partial_{j}
\partial_{i} Q^{(0)} + (1/3) \delta_{ij} Q^{(0)}$.
These expansions involve the independent functions $h$, $\eta$, $\delta $, $\theta$, 
$\sigma $, and $\Xi $, which depends on $k$ and $\tau$. 
For adiabatic perturbations, functions $\pi_{L} $ and $\delta $ are not independent since 
they must satisfy the 
relation $\pi_{L} = (\rho_{B} / p_{B})
(dp_{B}/d\rho_{B}) \delta$. 

Finally, quantities $\delta $, $\theta $ and $\sigma$ may be calculated 
by using the following formulas \citep{mb95}
\begin{equation}
\rho_{B} \delta = \sum_{i} \rho_{Bi} \delta_{i} \ ,
\end{equation}
\begin{equation}
(\rho_{B}+p_{B}) \theta = \sum_{i} (\rho_{Bi}+p_{Bi}) \theta_{i}  \ ,
\end{equation}
\begin{equation}
(\rho_{B}+p_{B}) \sigma = \sum_{i} (\rho_{Bi}+p_{Bi}) \sigma_{i}   \ ,
\end{equation}
where the subscript $i$ run over the particle species ($b$, $c$, $\nu$, $\gamma$).

Let us now summarize the evolution equations of the above scalar modes, which 
were derived in \cite{mb95,dal12,dal14} by using the above expansions and the 
VT equations.

The $\Xi$ evolution is governed 
by a second order differential equation \citep{dal12,dal14}, which is equivalent to the following
system of two first order differential equations:
\begin{equation}
\dot{\Xi}  = \xi
\label{fe12-l1}
\end{equation}    
\begin{equation}
\dot{\xi} = -2 \frac {\dot{a}}{a} \xi - k^{2} \Xi \ ;
\label{fe12-l2}
\end{equation}  
these equations do not involve the modes of GR, but only the VT quantity $\xi$,
the wavenumber,  
and quantities related to the background. This fact is very 
advantageous eqs.~(\ref{fe12-l1}) and~(\ref{fe12-l2}) are included into CAMB for
adaptation to VT estimates.

In the chosen gauge, the following 
linearized equations are also satisfied:
\begin{equation}
k^{2} \eta - \frac {1}{2} \frac {\dot{a}}{a} \dot{h} = 4 \pi G 
[-a^{2} \rho_{B} \delta - 2 \gamma \Xi_{B} (a^{2}
\Xi_{B} \Xi + A_{0B} \xi^{(0)} ) ]  
\label{mbl1}
\end{equation}  
\begin{equation}
k^{2} \dot{\eta} = 4 \pi G 
[a^{2} (\rho_{B}+p_{B}) \theta + 2 \gamma k^{2} A_{0B} \Xi_{B} 
\Xi]
\label{mbl2} 
\end{equation}  
\begin{equation}   
\ddot{h} + 2 \frac {\dot{a}}{a} \dot{h} -2k^{2} \eta = 
-24 \pi G 
[a^{2} p_{B} \pi_{L} - 2 \gamma \Xi_{B} (a^{2}
\Xi_{B} \Xi - A_{0B} \xi )]
\label{mbl3} 
\end{equation}  
\begin{equation}
\ddot{h} + 6 \ddot{\eta} + 2 \frac {\dot{a}}{a} (\dot{h} + 6\dot{\eta}) 
-2k^{2} \eta = 
-24 \pi G a^{2} (\rho_{B}+p_{B}) \sigma   \ .
\label{mbl4} 
\end{equation}    
The terms involving $\gamma $ are the VT corrections to the standard GR 
equations (21a)--(21d) derived in \cite{mb95}, which are formally recovered for $\gamma =0$. 
These terms       
--appearing only in VT cosmology-- have been  
included in CAMB; they have been proved to be independent of both $S_{gn} $
and $\gamma $.

Since the energy-momentum conservation law 
$\nabla_{\alpha} T^{\alpha \beta}_{GR} = 0$ is satisfied (as in GR),
the variables $\delta_{\gamma} $, $\delta_{\nu} $, $\theta_{\gamma} $, $\theta_{\nu } $
and $\sigma_{\nu}  $ obey the same equations as in GR
cosmology and, consequently, we can write (see eqs.~(92) in paper \cite{mb95}):
\begin{eqnarray}
& &
\dot{\delta}_{\gamma} +\frac {4}{3} \theta_{\gamma} +\frac{2}{3} \dot{h} =0, 
\,\,\,\, \dot{\theta}_{\gamma} - \frac {1}{4} k^{2} \delta_{\gamma} =0, \,\,\,\,  
\nonumber \\
& & 
\dot{\delta}_{\nu} +\frac {4}{3} \theta_{\nu} +\frac{2}{3} \dot{h} =0, 
\,\,\,\, \dot{\theta}_{\nu} - \frac {1}{4} k^{2} (\delta_{\nu} -4 \sigma_{\nu}) =0, \,\,\,\,                   
\nonumber \\
& &                                     
\dot{\sigma}_{\nu} - \frac {2}{15} (2 \theta_{\nu} + \dot{h} + 6 \dot{\eta}) =0 \ .
\label{arr2}
\end{eqnarray}
The Thompson interaction between photons 
and electrons (including reionization) is not affected by the presence 
of the vector field $A^{\mu} $ and, consequently, the CAMB 
treatment of this scattering is not to be modified.

Finally, as it was proved in paper I, the initial conditions for the VT scalar modes,
at redshift $z_{in} =10^{8} $ (radiation dominated era), are the following:
\begin{eqnarray}
& &
h = C (k\tau )^{2} + \tilde{C} (k\tau )^{4}, \,\,
\delta_{\gamma} = \delta_{\nu} = \frac {4}{3} \delta_{b} = \frac {4}{3} \delta_{c}
= -\frac {2}{3} h \ ,     
\nonumber \\
& &
\theta_{c} =0, \,\,\, \theta_{\gamma} = \theta_{b} = -\frac {1}{18} C k^{4} \tau^{3}
-\frac {1}{30} \tilde{C} k^{6} \tau^{5} \ ,
\nonumber \\
& &
\theta_{\nu} = - \frac {23+4R_{\nu}}{18(15+4R_{\nu})}
C k^{4} \tau^{3}-\frac{1}{30} \tilde{C} k^{6} \tau^{5} \ ,
\nonumber \\
& & 
\sigma_{\nu} = \frac {4}{3(15+4R_{\nu})} C k^{2} \tau^{2}, \,\,\,\, 
\Xi = D k^{4}, 
\,\,\, \xi = \dot{\Xi} = 0 \ ,   
\nonumber \\ 
& &
\eta = \Big[2- \frac {5+4R_{\nu}} {6(15+4R_{\nu})} (k\tau)^{2} \Big] C \ ,
\label{arr3}   
\end{eqnarray}  
where $D= 3(1+z_{in})^{2}\tilde{C}/[8\pi G\rho_{V} (\dot{a}/a)^{2}_{in}]$; hence,
these formulas involve two independent normalization constants $C$ and $D$ (or $C$ and $\tilde{C} $) and 
the quantity $R_{\nu} = \rho_{\nu B}/(\rho_{\nu B}+\rho_{\gamma B})$. 
Constant $C$ must be fixed as in GR to guarantee that VT perturbations reduce to those of GR as 
$D $ (constant $D_{1}$ in paper I) tends to zero.

In the study performed in paper I, which was based on the differential equations
and initial conditions summarised in this section, the following important numerical results 
were found:

1) For spatial scales $L \gtrsim 2800 h^{-1} \ Mpc$, the VT quantities
$\dot{h} $ and $\dot{\eta} $ are almost identical to those of GR for any redshift $z$; 
however, for $L \lesssim 2800 h^{-1} \ Mpc$, the VT values of $\dot{h} $ and $\dot{\eta} $
deviate significantly from the GR values for $z \lesssim 5$. The resulting deviations 
involve oscillations. 
Since quantities $\dot{h} $ and $\dot{\eta} $ are explicitly involved in the 
equations describing the evolution of the photon distribution function 
(see eq.~(63) in \cite{mb95}), their deviations -with respect to GR- must leave 
imprints on the CMB temperature angular power spectrum; however, 
these deviations should not affect CMB polarization, which is essentially generated 
during recombination ($z \sim 1100$). This fact was numerically verified 
in paper I.

2) For $\ell \gtrsim 250$ ($\ell \lesssim 5$) the $C_{\ell} $ multipoles of GR and VT are 
almost identical (very similar);
hence, the angular power spectrum of VT only deviates -significantly- with respect to that of GR for
$5 \lesssim \ell \lesssim 250$. The deviations depend on $|D| $, but they are independent of the sign of $D$.

As it is pointed out below, these results have been very useful to properly modify 
the codes CAMB and COSMOMC \citep{lew02} for VT applications.

\subsection{On the codes VT-CAMB and VT-COSMOMC}  
\label{sec:2-1}          

It is worthwhile to describe some changes, which have been necessary to built up VT-CAMB starting from CAMB. 
All these changes are suggested by the outcomes summarized in 
section~\ref{sec:2}. The basic VT cosmological equations and initial conditions
-at $z_{in} =10^{8} $- must be implemented in VT-CAMB as it has been discussed in 
the aforementioned section; nevertheless, other changes -in CAMB- are also necessary 
to ensure high enough accuracy in VT-CAMB predictions. 
The most important of these changes is now described.

The code CAMB estimates some integrals along the background null geodesics -until 
vanishing redshift- to get the CMB temperature and polarization anisotropy. 
Since it has been emphasized in section~\ref{sec:2} that VT and GR are 
almost equivalent at $z > 5$, and also that the deviations arising at 
$z \lesssim 5 $ involve oscillations. It is evident that many integration steps must be 
included in the redshift interval (0,$\sim$5) to properly take into account the 
deviations between VT and GR. We have estimated the minimum number of
integration steps leading to satisfactory results. 
More steps are not necessary since they do not 
improve on the results. In the original CAMB code, the number of steps in the 
interval (0,$\sim$5) is not sufficient for VT applications.  

The COSMOMC version we have modified uses the so-called {\em nuisance} parameters,
which have been defined to take into account -in the context of Planck experiment- 
contaminant foregrounds, beam structure, and so on; hence, 
the treatment of Planck data cannot be realized with COSMOMC versions designed to 
deal only with WMAP data (old versions as that modified in paper I).
Modifications necessary to get VT-COSMOMC are simple; this code calls VT-CAMB and includes the new 
parameter $D$ to be adjusted.

If the code VT-CAMB is used for $D=0$, results are not identical to 
those of the original CAMB, which is due to the fact that, even for $D=0$,
the number of integration time steps used by CAMB and VT-CAMB are very different.
It has been verified that the differences between these two codes are negligible
for $D=0$, which means that the original CAMB code is accurate enough 
for the standard GR cosmological model (hereafter GR-CM), and no more time steps are necessary.

\section{Estimating the parameters of the VT cosmological mode}  
\label{sec:3}

In general, parameter estimates require: (i) accurate enough observational data for an appropriate set of 
observable quantities, (ii) a numerical code predicting the values of these observable 
quantities for given values of appropriate cosmological parameters; e.g.,  
CAMB is a code of this kind, which has been designed to work in the 
framework of GR-CM, and 
(iii) another numerical code based on suitable 
statistical methods as, e.g., Markov chains, which allows us to fit 
current observational data and numerical predictions. A code of this type 
is COSMOMC, which was designed to work in the GR-CM.

\subsection{On the cosmological parameters}  
\label{sec:3-1}          

The following basic assumptions are maintained all along the paper:
the background universe is flat, perturbations are adiabatic, the dark energy equation of state is
$p = W \rho $ with $W=-1$, vector modes are negligible,
the mean CMB temperature is $T_{CMB}=2.726$, and the effective number of relativistic species is 
$3.046$.

At horizon crossing, the power spectrum of the scalar energy density perturbations is 
parametrized as follows:
\begin{equation} 
P_{s}(k) = A_{s} \Big(\frac {k}{k_{0}}\Big)^{n_{s}-1+(1/2)(dn_{s}/d\,ln\,k)ln(k/k_{0})} \ ,
\label{eps}
\end{equation}  
whereas the power spectrum of the gravitational wave amplitudes is of the form 
\begin{equation}  
P_{t}(k) = A_{t} \Big(\frac {k}{k_{0}}\Big)^{n_{t}}\ .
\label{ept}
\end{equation}  
The pivot scale is
$k_{0} = 0.05 \ Mpc^{-1}$. It is usual to define the parameter 
$r_{0.05} = A_{t}/A_{s} $, which depends on $n_{t} $ in most relevant cases; e.g.,  
in inflationary models based on a scalar field, the so-called consistency condition
for slow-roll inflation,  $r_{0.05} = -n_{T}/8 $, is satisfied.
Since this condition is assumed here, as it is done in \cite{pla14a}, the free independent parameters involved in 
eqs.~(\ref{eps})-(\ref{ept}) are $A_{s}$, $r_{0.05} $, $n_{s} $, and $dn_{s}/d\,ln\,k$, which 
are the normalization constant of $P_{s}(k)$, the primordial tensor to scalar ratio, 
the spectral index of $P_{s}(k)$, and the running index, respectively.
Sometimes, for comparisons with \cite{pla14a}, we use the parameter $r_{0.002}$,
which is defined as $r_{0.05}$, but assuming the pivot scale 
$k_{0} = 0.002 \ Mpc^{-1}$.

The six parameters 
used to fit predictions and observations in the GR-CM (minimal fit model) are
$\Omega_{b}h^{2}$, $\Omega_{c}h^{2} $, $\tau $, $n_{s} $,
$\log[10^{10}A_{s}]$, and $\theta_{MC} $, where 
$\Omega_{b}$ and $\Omega_{c}$ 
are the density parameters of baryons and dark matter, respectively,
$h$ is the reduced Hubble constant, $\tau $ is the reionization optical depth, and
the parameter $\theta_{MC} $ (angular acoustic scale) is the ratio
$r_{s}(z_{*})/d_{A}(z_{*})$, where 
$r_{s}(z_{*})$ is the sound horizon at decoupling redshift $z_{*} $ and
$d_{A}(z_{*}) $ is the angular diameter distance
for the same redshift.

All the parameters involved in standard GR cosmology are also 
parameters of the VT cosmological model (hereafter VT-CM); nevertheless, in the VT case, 
there is an additional parameter denoted $D$ (see section~\ref{sec:2}). 

The minimal fit model (for VT-CM), used here and also in paper I, involves the six parameters of the minimal GR-CM fit
plus $D$.

\subsection{Fit models}  
\label{sec:3-2}          

Various fits have been considered -in this paper- to study the 
VT-CM viability. They improve on the fit used in paper I to 
estimate cosmological parameters in VT-CM.
The main differences between the five fit models of this paper 
and the fit approach of paper I are now pointed out.

The fit method of paper I was designed as follows:
(a) the seven cosmological parameters of the minimal fit (see above) were 
assumed,
(b) appropriate versions of codes CMBFAST \citep{seza96} and COSMOMC (january 2012 version) 
were modified and coupled with the essential aim of estimating the seven chosen parameters, 
(c) vector and tensor modes were not considered at all to do predictions, and 
(d) only data about Ia supernovae (SNIa) and WMAP7 CMB anisotropies were taken into account.

For comparisons,
let us describe the five fit models 
analyzed in this paper. They are all based on the use of 
the VT-CAMB and VT-COSMOMC
numerical codes, which are adaptations of original CAMB and COSMOMC
versions (december 2013), specially designed -by us- to be applied 
in the case $D \neq 0$. 

Let us now focus our attention on the features of the five fit models we have selected:

In the first fit, we use the same cosmological parameters and perturbation 
modes as in the approach of paper I (see above); however, the updated (december 2013)
data sets are Planck CMB anisotropies (Planck), and WMAP polarization anisotropy at 
low $\ell \lesssim 23 $ (WP). Following a notation similar to that of Planck papers \citep{pla14a}, this fit is 
hereafter named Planck+WP

Our second (third) fit is like Planck+WP, but it also consider updated BAO (SNIa)
information; hence, it is named Planck+WP+BAO (Planck+WP+SNIa)

The fourth fit model includes the same data sets as Planck+WP, plus primordial gravitational 
waves (tensor modes $\equiv$ TM). The additional parameter
$r_{0.05} $ is then necessary. This model is hereafter called Planck+WP-TM

Finally, the fifth fit is like Planck+WP-TM, but 
an additional parameter $dn_{s}/dlnk$ is included to analyze the effects of a running 
spectral index (RSI) with a weak dependence on $k$. This fit is named
Planck+WP-TM-RSI.

Since we are particularly interested in possible differences between GR-CM and VT-CM,
parameters related with particle interactions as, e.g., the total neutrino mass $\sum m_{\nu}$
(summed over the three neutrino families),
the effective neutrino number $N_{eff}$ (relativistic particles), and so on (see \cite{kot94,pla14a,cos14}), 
are not considered in our fits. This procedure is qualitatively justified by the fact that
the CMB anisotropy due to physical interactions among particles must be produced inside the effective horizon
at $z \gtrsim 1100$, namely, at angular scales corresponding to $\ell \gtrsim 220$ (in a flat universe), 
which means that no important effects 
are expected at $\ell \lesssim 100 $; namely, at the angular scales producing the most important 
deviations between VT and GR. As an additional result of the these considerations, high $\ell $ data from
SPT (South Pole Telescope) and ACT (Atacama Cosmology Telescope) 
small scale CMB experiments should not produce relevant differences 
between GR-CM and VT-CM and, consequently, this high $\ell $ information is not 
considered in our fits.

\section{Results}
\label{sec:4}

A systematic comparison of the parameters obtained
for $D = 0$ (GR-CM) and $D \neq 0$ (VT-CM) has been performed
for the five fit models described above, in each of them, codes VT-CAMB and VT-COSMOMC are used 
both for $D = 0$
and for $D \neq 0$. Whatever the fit may be, parameters
different from $D$ and observational data are the same for both $D=0$ and $D\neq 0$; 
in this way, differences between GR-CM ($\Lambda$CDM) and VT-CM
are properly estimated.

Planck collaboration has developed an exhaustive study to estimate
the cosmological parameters -by using CAMB and COSMOMC- 
in the context of GR-CM \citep{pla14a}. Some feasible comparisons between our 
VT-CM results and those of the Planck team (GR-CM) are presented.

Let us now present our results and comparisons for the chosen fit models.

\subsection{Planck+WP }
\label{sec:4-1}

\begin{figure*}                   
\centering
\includegraphics[width=16.cm]{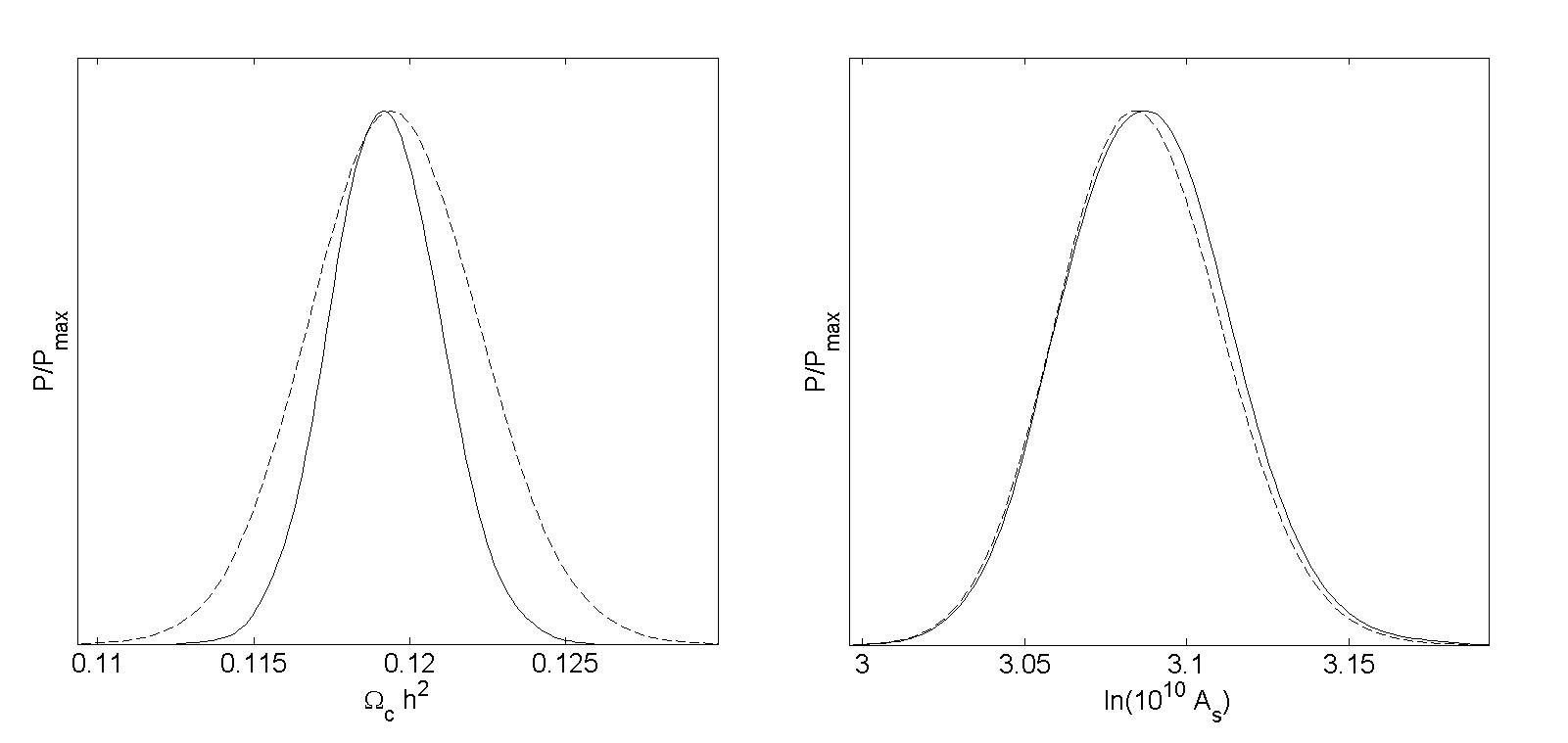}  
\caption{Marginalized distribution functions normalized to unity for the parameters $\Omega_{c}h^{2}$ (left) 
and $\ln{(10^{10}A_{s})}$ (right). Continuous (dashed) lines correspond to GR (VT)} 
\label{figu1}%
\end{figure*}

\begin{figure*}                   
\centering
\includegraphics[width=16.cm]{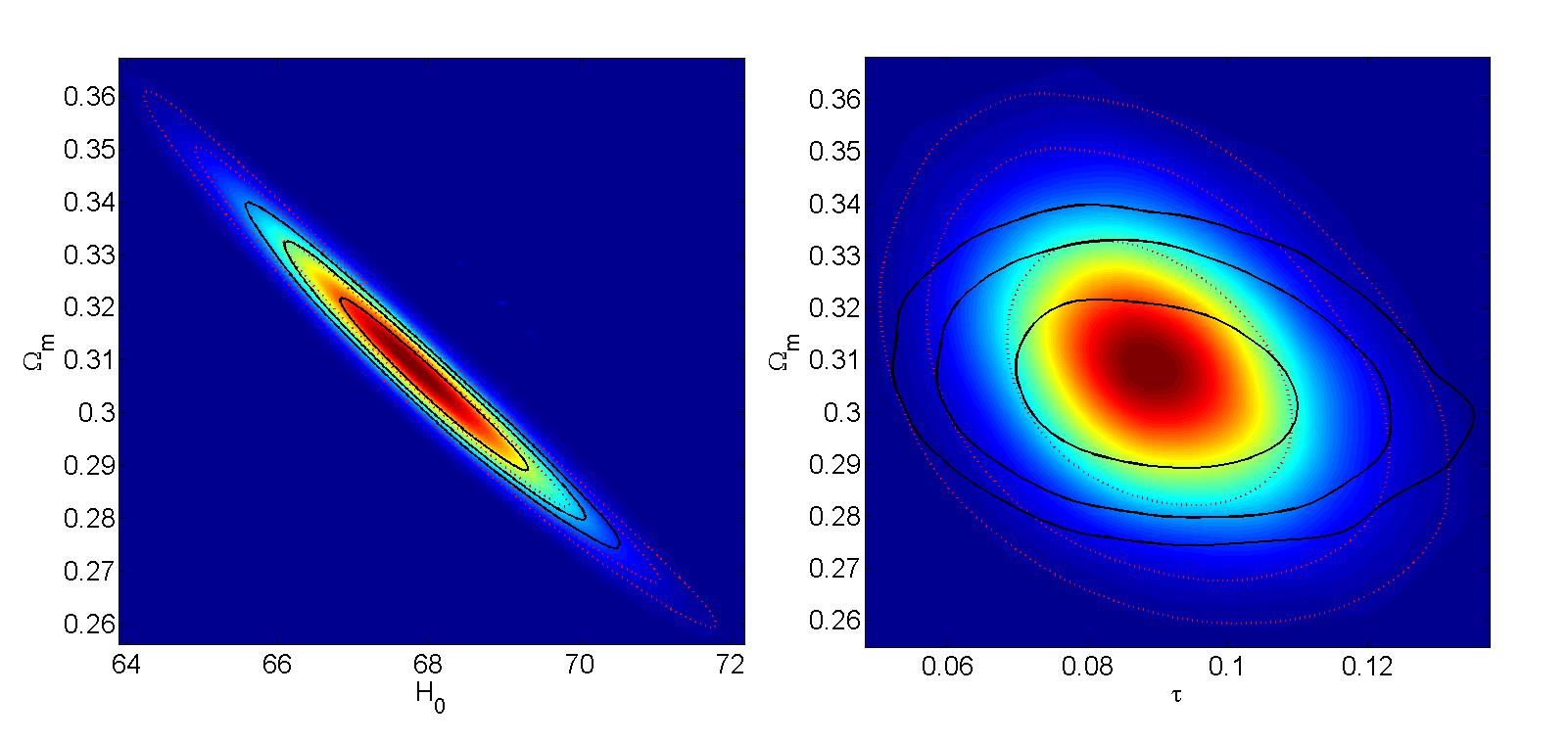}
\caption{Marginalized distribution functions (color), for the pairs ($H_{0}$, $\Omega_{m}$) [left]
and ($\tau$, $\Omega_{m}$) [right]. Red [black] contours correspond to the $1\sigma$ (inner),
$2\sigma$ (middle) and $3\sigma$ (outer) confidence levels in VT [GR]. 
} 
\label{figu2}%
\end{figure*}

Results corresponding to this fit, for $D=0$ (GR) 
and for $D \neq 0$, are compared in Table~\ref{table:1} and figures~\ref{figu1} and~\ref{figu2}. 

The first column shows fourteen parameters. The first seven of them (above the horizontal line)
are the fitted parameters in VT, and the remaning ones are given by CAMB as 
derived parameters. The second, third and fourth columns display, in VT, 
the best fit (BF) values, the lower (L1) limit at 
$1\sigma$ confidence level, and the corresponding upper (U1) limit, respectively. 
For comparison with the results 
obtained in GR, the relative deviations $\Delta(L1)=2[L1(VT)-L1(GR)]/[L1(VT)+L1(GR)]$ 
and $\Delta(U1)=2[U1(VT)-U1(GR)]/[U1(VT)+U1(GR)]$ are presented in the 
fifth and sixth columns, respectively. Finally the seventh column gives 
the ratio $R1=[U1(VT)-L1(VT)]/[U1(GR)-L1(GR)]$ between the amplitudes of the 
$(U1,L1)$ intervals of VT and GR. All the Tables presented below have 
very similar structures, with small changes to be described in due time.

\begin{table*}
\caption{Planck+WP fit}             
\label{table:1}      
\centering          
\begin{tabular}{l l l l l l l }     
\hline\hline       
Parameter & BF & L1 & U1 & $\Delta(L1)$ & $\Delta(U1)$ & R1\\
\hline                    
$D \times 10^{-8}$ & 1.596 & 0.000 & 2.149 & - & - & - \\  
$\Omega_{b}h^{2}$ & 0.02216 & 0.02179 & 0.02235 & -0.27\% & 0.04\% & 1.143 \\
$\Omega_{c}h^{2}$ & 0.1187 & 0.1169 & 0.1222 & -0.51\% & 0.99\% & 1.514\\
$100\theta_{MC}$ & 1.0411 & 1.0407 & 1.0419& -0.01\% & 0.00\% & 1.091 \\
$\ln{(10^{10}A_{s})}$ & 3.085 & 3.060 & 3.110 & -0.10\% & -0.10\% & 1.000\\
$n_{s} $ & 0.9657 & 0.9535 & 0.9684 & -0.18\% & 0.20\% & 1.319 \\
$\tau $ & 0.0893 & 0.0749 & 0.1013 & -1.46\% & -0.88\% & 1.008 \\
\hline
$\Omega_{\Lambda} $ & 0.697 & 0.677 & 0.710 & -1.12\% & 0.59\% & 1.557 \\
$\Omega_{m} $ & 0.303 & 0.290 & 0.323& -1.44\% & 2.38\% & 1.557 \\
$\sigma_{8} $ & 0.838 & 0.827 & 0.853 & -0.10\% & 0.13\% & 1.082 \\
$z_{re} $ & 10.98 & 9.86 & 12.04 & -0.81\% & -0.66\% & 1.000 \\
$H_{0} $ & 68.22 & 66.72 & 69.12 & -0.82\% & 0.41\% & 1.529 \\
$Y_{P} $ & 0.24488 & 0.24473 & 0.24496 & -0.01\% & 0.00\% & 1.137 \\
$t_{0} $ & 13.78 & 13.74 & 13.83 & 0.00\% & 0.07\% & 1.125 \\
\hline                  
\end{tabular}
\end{table*}

Since we have verified that results do not depend on the sign of $D$ 
(see also paper I),  
from Table~\ref{table:1}, it follows that, for positive (negative) values of 
$D \times 10^{-8} $, this quantity belongs to the interval [0,2.149] ([-2.149,0]) 
with a probability $\sim 68.2 \%$ ($1\sigma $). Parameter $D$ -characteristic of VT-
is also adjusted in other fits (see below).
Results for different fits will be compared below to discuss the 
statistical role of $|D|$.
                                     
In the fifth and sixth columns, we see that the relative deviations 
-between VT and GR- measured by  $\Delta L1$ and $\Delta U1$ are small for all 
the adjusted parameters. Parameters $\tau $ and $\Omega_{c}h^{2}$ 
undergo the maximum relative deviations, which are small in both 
cases since $|\Delta L1|$ and $|\Delta U1|$ do not exceed $1.46\%$.
Finally, the seventh column shows that, at $1\sigma$ level, 
the inequality $R1 \geq 1$ is satisfied for the forteen parameters,
which means that, for every parameter, the amplitude of the interval 
$(U1,L1)$ in VT is greater than in GR. This strongly suggests that 
a parameter $D \neq 0$ facilitates the adjustements between predictions 
and cosmological observations. 

The same can be seen in figure~\ref{figu1},
where the marginalized likelihood function $P/P_{max}$ corresponding 
to VT (dashed line) is wider than that of GR (solid line) 
for the parameter $\Omega_{c}h^{2}$ ($R\simeq 1.5$ in Table 1); however, 
for $\ln{(10^{10}A_{s})}$, having $R1 \simeq 1$, both 
likelihood functions are almost identical as it was expected.
Also figure~(\ref{figu2}) displays the same situation; in fact, from inside out, 
red (black) curves show the $1\sigma$, $2\sigma$
and $3\sigma$ limits for VT (GR). Moreover, in the left panel 
[pair ($H_{0}$, $\Omega_{m}$)] as well as in the right panel
[pair ($\tau$, $\Omega_{m}$)], we see that -almost everywhere- the red curves are 
outside the corresponding black lines.

\subsection{Planck+WP+BAO }
\label{sec:4-2}

\begin{figure*}                   
\centering
\includegraphics[width=16.cm]{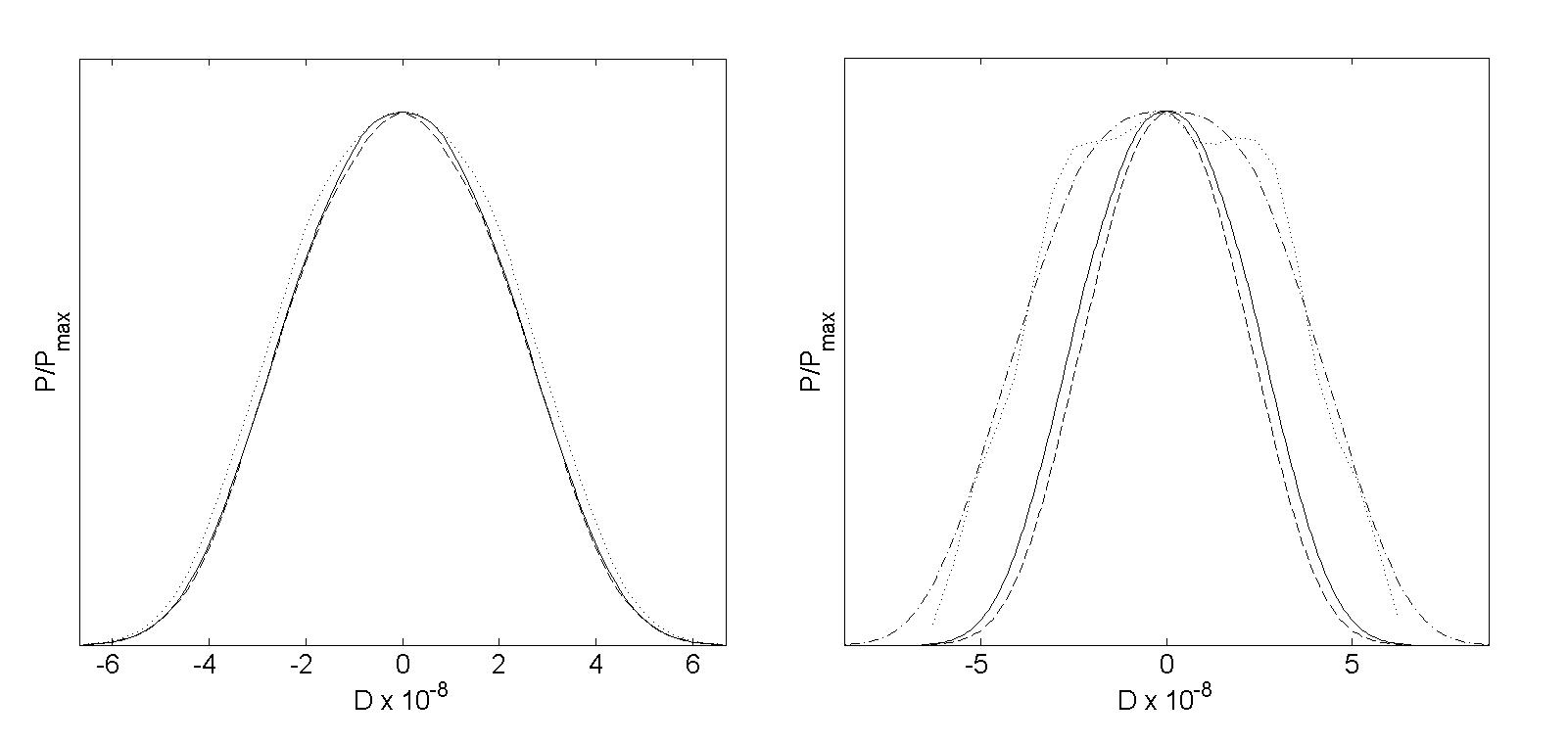}
\caption{Marginalized distribution functions normalized to unity ($P/P_{max}$) for the parameter 
$D\times 10^{-8}$ in various fits. Left: solid, dashed, and dotted lines correspond to the Planck-WP, Planck-WP+BAO,
and Planck-WP-SNIa, respectively. Right: pointed curve was obtained with WMAP and SNIa data 
in paper I, whereas solid, dashed, and dot-dashed lines give $P/P_{max}$ in Planck-WP, 
Planck+WP-TM, and Planck+WP-TM-RSI, respectively
} 
\label{figu3}%
\end{figure*}

In this fit. Results for $D=0$ (GR) 
and for $D\neq 0$ (VT) are compared in Table~\ref{table:2} and 
figure~\ref{figu3}. From the Table it follows that,
for all the parameters, $|\Delta L1|$ and $|\Delta U1|$ 
do not exceed $0.92\%$ and $R1$ is very close to unity. If these results are 
compared with those of the Planck+WP fit of section~\ref{sec:4-1} 
(Table~\ref{table:1}), we see that BAO data have reduced the differences 
between GR and VT at the $1\sigma$ level for all the common parameters.
From the left panel of figure~\ref{figu3}, 
it follows that the solid (Planck+WP) and dashed  (Planck+WP+BAO) lines
are almost identical for $D \times 10^{-8}\lesssim 0.4 $ and for $D\times 10^{-8} \gtrsim 1.7 $, which 
clearly explains that the best fit value of $D\times 10^{-8}$ is as small as 
0.316, and also that
the $1\sigma$ limit of parameter $D\times 10^{-8}$ takes on the value $2.149$, which is identical
to that of the Planck+WP case. The $1\sigma$ limit is a meaningful 
quantity; however, as it is commented in \cite{pla14a} (last paragraph of section 2), 
best fit values are not very numerically stable and should not be over-interpreted.
The probabilities assigned in COSMOMC have numerical errors and, consequently,  
inside a flat enough region of the $D$ distribution function, the 
maximum likelihood value of $D$ could arise as a result of
these errors; which makes this value unstable against the number of 
selected chains (convergence criterium). On account of these comments,
best fits are hereafter interpreted with caution.

\begin{table*}
\caption{Planck+WP+BAO fit}             
\label{table:2}      
\centering          
\begin{tabular}{l l l l l l l }     
\hline\hline       
Parameter & BF & L1 & U1 & $\Delta(L1)$ & $\Delta(U1)$ & R1\\
\hline                    
$D \times 10^{-8}$ & 0.316 & 0.000 & 2.149 & - & - & - \\  
$\Omega_{b}h^{2}$ & 0.02205 & 0.02184 & 0.02233 & -0.05\% & -0.04\% & 1.000 \\
$\Omega_{c}h^{2}$ & 0.1190 & 0.1174 & 0.1209 & -0.09\% & 0.00\% & 1.029\\
$100\theta_{MC}$ & 1.0412 & 1.0408 & 1.0419& 0.00\% & 0.00\% & 1.000 \\
$\ln{(10^{10}A_{s})}$ & 3.067 & 3.060 & 3.110 & -0.07\% & -0.10\% & 0.980\\
$n_{s} $ & 0.9602 & 0.9559 & 0.9676 & 0.06\% & 0.09\% & 1.026 \\
$\tau $ & 0.0795 & 0.0758 & 0.1016 & -0.92\% & -0.88\% & 0.992 \\
\hline
$\Omega_{\Lambda} $ & 0.695 & 0.685 & 0.705 & 0.01\% & 0.07\% & 1.020 \\
$\Omega_{m} $ & 0.305 & 0.295 & 0.315& -0.17\% & -0.03\% & 1.020 \\
$\sigma_{8} $ & 0.831 & 0.827 & 0.851 & -0.07\% & -0.09\% & 0.992 \\
$z_{re} $ & 10.15 & 9.92 & 12.07 & -0.60\% & -0.74\% & 0.986 \\
$H_{0} $ & 68.02 & 67.31 & 68.87 & 0.01\% & 0.06\% & 1.020 \\
$Y_{P} $ & 0.24484 & 0.24475 & 0.24496 & 0.00\% & 0.00\% & 1.000 \\
$t_{0} $ & 13.79 & 13.74 & 13.82 & 0.00\% & 0.00\% & 1.000 \\
\hline                  
\end{tabular}
\end{table*}

\subsection{Planck+WP+SNIa }
\label{sec:4-3}

For this third fit, results corresponding to $D=0$ (GR) 
and $D\neq 0$ (VT) may be compared with the help of Table~\ref{table:3} and figure~\ref{figu3}. In this Table,
one can see that, for all the parameters, quantities $|\Delta L1|$ and $|\Delta U1|$ 
do not exceed $0.5\%$ and $R$ is very close to unity. If these results are 
compared with those of the Planck+WP fit of section~\ref{sec:4-1} 
(Table~\ref{table:1}), we see that, as it occurs with BAO,  SNIa data have also reduced the differences 
between GR and VT, at the $1\sigma$ level, for all the parameters being common to both theories.
For positive values,               
the $1\sigma$ limit of $D\times 10^{-8}$ takes on the value $2.245$, which is a little greater 
than that of the Planck+WP case ($2.149$). 
This is in agreement with the fact that, in the left panel of figure~\ref{figu3}, 
the dotted (Planck+WP+SNIa) line is a little wider that the solid (Planck+WP) curve.
The best fit value of $D\times 10^{-8}$ is 0.756, which is located in the flat central part of 
the dotted curve of figure~\ref{figu3}; hence, this value is little meaningful 
(see section~\ref{sec:4-2}).

\begin{table*}
\caption{Planck+WP+SNIa fit}             
\label{table:3}      
\centering          
\begin{tabular}{l l l l l l l }     
\hline\hline       
Parameter & BF & L1 & U1 & $\Delta(L1)$ & $\Delta(U1)$ & R1\\
\hline                    
$D \times 10^{-8}$ & 0.756 & 0.000 & 2.245 & - & - & - \\  
$\Omega_{b}h^{2}$ & 0.02208 & 0.02191 & 0.02246 & 0.03\% & 0.06\% & 1.012 \\
$\Omega_{c}h^{2}$ & 0.1200 & 0.1154 & 0.1203 & -0.21\% & -0.21\% & 0.997\\
$100\theta_{MC}$ & 1.0414 & 1.0409 & 1.0421& 0.00\% & 0.00\% & 1.000 \\
$\ln{(10^{10}A_{s})}$ & 3.097 & 3.060 & 3.125 & -0.05\% & -0.06\% & 0.994\\
$n_{s} $ & 0.9630 & 0.9577 & 0.9720 & 0.13\% & 0.16\% & 1.016 \\
$\tau $ & 0.0930 & 0.0772 & 0.1043 & -0.37\% & -0.49\% & 0.991 \\
\hline
$\Omega_{\Lambda} $ & 0.691 & 0.688 & 0.717 & 0.23\% & 0.19\% & 0.993 \\
$\Omega_{m} $ & 0.309 & 0.283 & 0.312& -0.48\% & -0.50\% & 0.993 \\
$\sigma_{8} $ & 0.847 & 0.823 & 0.848 & -0.13\% & -0.13\% & 0.998 \\
$z_{re} $ & 11.36 & 10.02 & 12.22 & -0.33\% & -0.35\% & 0.990 \\
$H_{0} $ & 67.78 & 67.55 & 69.82 & 0.17\% & 0.15\% & 0.996 \\
$Y_{P} $ & 0.24485 & 0.24477 & 0.24501 & 0.00\% & 0.00\% & 1.043 \\
$t_{0} $ & 13.79 & 13.71 & 13.81 & -0.07\% & 0.00\% & 1.111 \\
\hline                  
\end{tabular}
\end{table*}

\subsection{Planck+WP-TM }
\label{sec:4-4}

The three VT fits considered in previous sections are minimal (seven parameters).
Although minimal fits in GR-CM (six parameters) have led to very good results in the analysis of WMAP 
\citep{jar11,hin12} and Planck \citep{pla14a,pla14b,pla15a,pla15b} data; extended 
fits with more parameters have been also considered \citep{pla14a,div15}.
Here, and also in next section, 
new parameters are introduced with the essential aim of analyzing physically 
relevant problems. Since cosmic gravitational waves may significantly contribute 
to the CMB angular power spectrum for $\ell \lesssim 100 $, and 
the most important deviations between the VT and GR temperature multipoles
just arise for these $\ell $ values, 
some differences between the GR and VT parameters $r_{0.05}$ and $r_{0.002}$
seem to be possible and,
consequently, our attention is now focused on the Planck+WP-TM fit, which 
includes tensor modes.

\begin{figure*}                   
\centering
\includegraphics[width=15.8cm]{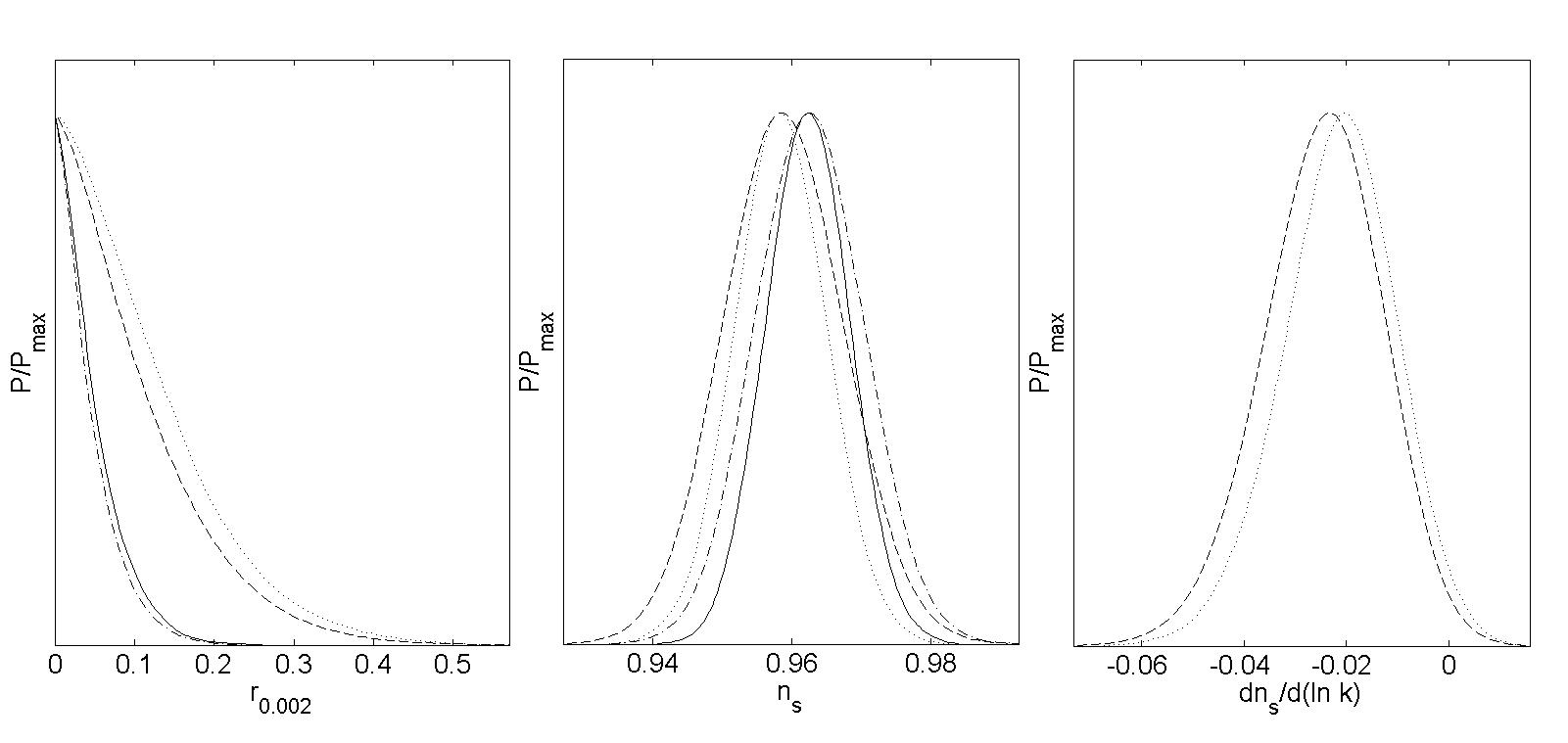}
\caption{Marginalized distribution functions normalized to unity.
In all panels, the solid and dot-dashed (dotted and dashed) lines correspond to the  
Planck-WP-TM (Planck-WP-TM-RSI) fit. Dashed and dot-dashed (dotted and solid) curves are
obtained in VT-CM (GR-CM). Each panel  
shows $P/P_{max}$ for the parameter specified below the horizontal axis.
} 
\label{figu4}%
\end{figure*}

\begin{figure*}                   
\centering
\includegraphics[width=16.cm]{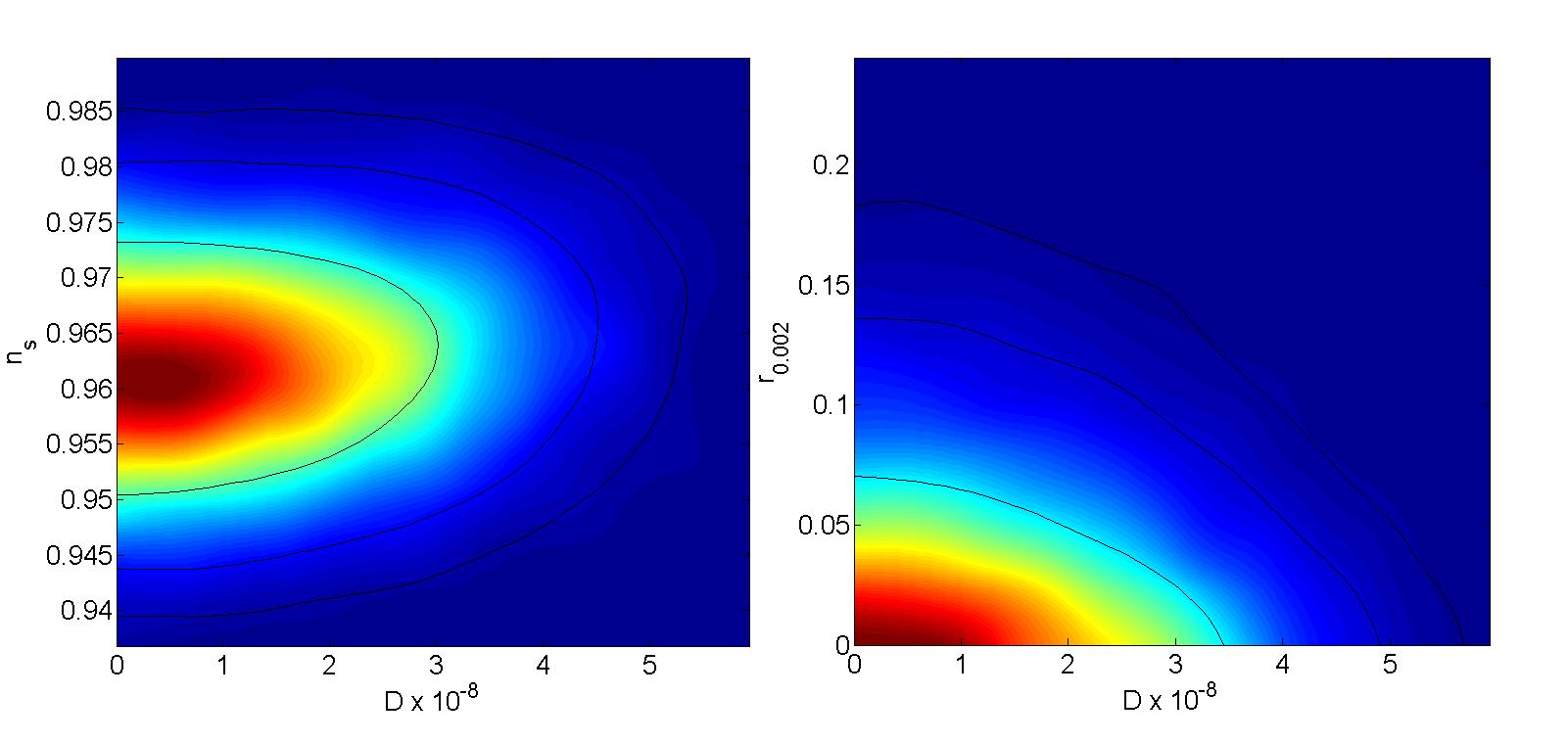}
\caption{Marginalized distribution functions (color), in the Planck+WP-TM fit. Two pairs involving $D\times 10^{-8}$, one with 
$n_{s}$ [left] and the other with $r_{0.002}$ [right], are considered. Contours correspond to the $1\sigma$ (inner),
$2\sigma$ (middle) and $3\sigma$ (outer) confidence levels in VT.
} 
\label{figu5}%
\end{figure*}

GR ($D=0$) and VT ($D \neq 0$) results may be compared by using Table~\ref{table:4} 
and figures~\ref{figu3}-\ref{figu5}. In this Table, columns 3-7 show quantities
as those of Tables~\ref{table:1}-~\ref{table:3}, but calculated at $2\sigma $
confidence level (probability around 95\%). This choice allow us to compare 
our results with those of \cite{pla14a}. 

In the fit of this section, which includes tensor modes, 
the $2\sigma $ upper limit of $D\times 10^{-8}$ -displayed in Table~\ref{table:4}-
is 3.665, whereas in the Planck-WP fit, the corresponding $2\sigma $ limit is 3.894; hence,
this limit is only weakly influenced by tensor modes. 
This is consistent with the right panel of figures~\ref{figu3}, where we see that the 
solid line (Planck+WP) is a little wider than the dashed curve (Planck+WP-TM).
Moreover, tensor modes have
reduced the best fit value of $D$ (compare Tables~\ref{table:1} and~\ref{table:4}),
although it is not highly significant.
Parameters $r_{0.05}$ (adjusted) and 
$r_{0.002}$ (derived) deserve attention. As it follows from the last row of Table~\ref{table:4},
the relative deviation -between VT and GR- corresponding to $r_{0.002}$ is $\Delta (U2) \simeq -6.20\%$
and the $2\sigma $ upper limit in VT is $U2 \simeq 0.1086$; hence, the $2\sigma $ upper limit in 
GR is found to be $U2 \simeq 0.11554$ and, consequently, we can write:

$r_{0.002} < 0.1086$ (VT, $\simeq 95\%$)

and 

$r_{0.002} < 0.11554$ (GR, $\simeq 95\%$).

This last inequality is to be compared with eq.~(63a) in \cite{pla14a}, where one can read
$r_{0.002} < 0.11$ ($\simeq 95\%$) for a Planck+WP+highL-TM fit -in GR- performed with the
original CAMB and COSMOMC codes. In practice, these 
two bounds are almost equivalent. The first of the above inequalities (VT) is also very similar to 
that of (GR). All this is in agreement with the right panel of figure~\ref{figu3}, where one may 
see that dashed curve (Planck+WP-TM) is located a little below the solid line (Planck+WP)
but very close to it.
The remaining adjusted parameters, common to the Planck+WP and Planck+WP-TM fits, do not
lead to remarkable news; e.g., the spectral index $n_{s} $ is considered in the 
central panel of figure~\ref{figu4}, where we see that
the dot-dashed line (Planck+WP-TM) is wider than 
the solid curve (Planck+WP), which is consistent with the value $R2=1.217$ displayed 
in Table~\ref{table:4}. We have focused our attention on $n_{s} $ since this parameter is 
in the exponent of eq.~(\ref{eps}) together with the running spectral index, 
which will be included 
in next fit. Finally, compare the solid (Planck+WP) and dot-dashed lines (Planck+WP-TM)
in the left panel of figure~\ref{figu4}] to see that the probability of any $r$ 
value is rather similar 
in GR and VT.

With the three basic parameters studied in the last paragraph, we have built up the pairs 
$(D\times 10^{-8}, n_{s})$ and $(D\times 10^{-8}, r_{0.002})$, whose marginalized distribution functions are
displayed in figure~\ref{figu5}. Since parameter $D$ vanishes in GR, only VT contours are 
shown (black curves). These contours confirm (see above) that, at $1\sigma$ level,
one satisfies $|D|\times 10^{-8} \lesssim 3$, showing also $D$ upper limits for $2\sigma$ 
and $3\sigma$ confidence levels.

\begin{table*}
\caption{Planck+WP-TM fit}             
\label{table:4}      
\centering          
\begin{tabular}{l l l l l l l }     
\hline\hline       
Parameter & BF & L2 & U2 & $\Delta(L2)$ & $\Delta(U2)$ & R2\\
\hline                    
$D \times 10^{-8}$ & 0.783 & 0.000 & 3.6655 & - & - & - \\  
$\Omega_{b}h^{2}$ & 0.02219 & 0.02154 & 0.02265 & -0.37\% & 0.27\% & 1.144 \\
$\Omega_{c}h^{2}$ & 0.1186 & 0.1139 & 0.1244 & -1.48\% & 1.62\% & 1.544\\
$100\theta_{MC}$ & 1.0417 & 1.0401 & 1.0429& -0.01\% & 0.04\% & 1.217 \\
$\ln{(10^{10}A_{s})}$ & 3.080 & 3.038 & 3.136 & -0.03\% & 0.00\% & 1.010\\
$n_{s} $ & 0.9629 & 0.9480 & 0.9777 & -0.32\% & 0.39\% & 1.297 \\
$\tau $ & 0.0847 & 0.0645 & 0.1157 & -1.54\% & 0.87\% & 1.041 \\
$r_{0.05}$ & 0.0325& 0.0000& 0.11588& - &-6.20\% & 0.940\\
\hline
$\Omega_{\Lambda} $ & 0.700 & 0.662 & 0.726 & -2.05\% & 1.39\% & 1.588 \\
$\Omega_{m} $ & 0.300 & 0.274 & 0.338& -3.58\% & 4.13\% & 1.588 \\
$\sigma_{8} $ & 0.835 & 0.814 & 0.864 & -0.32\% & 0.29\% & 1.112 \\
$z_{re} $ & 10.57 & 8.80 & 13.11 & -0.68\% & 0.23\% & 1.021 \\
$H_{0} $ & 68.48 & 65.73 & 70.54 & -1.36\% & 1.18\% & 1.562 \\
$Y_{P} $ & 0.24490 & 0.24461 & 0.24510 & -0.02\% & 0.01\% & 1.162 \\
$t_{0} $ & 13.76 & 13.68 & 13.88 & -0.22\% & 0.22\% & 1.429 \\
$r_{0.002}$ & 0.00289& 0.00000 &0.10859 & - & -6.20\%   0.940\\
\hline                  
\end{tabular}
\end{table*}

\subsection{Planck+WP-TM-RSI}
\label{sec:4-5}

\begin{figure*}                   
\centering
\includegraphics[width=16.cm]{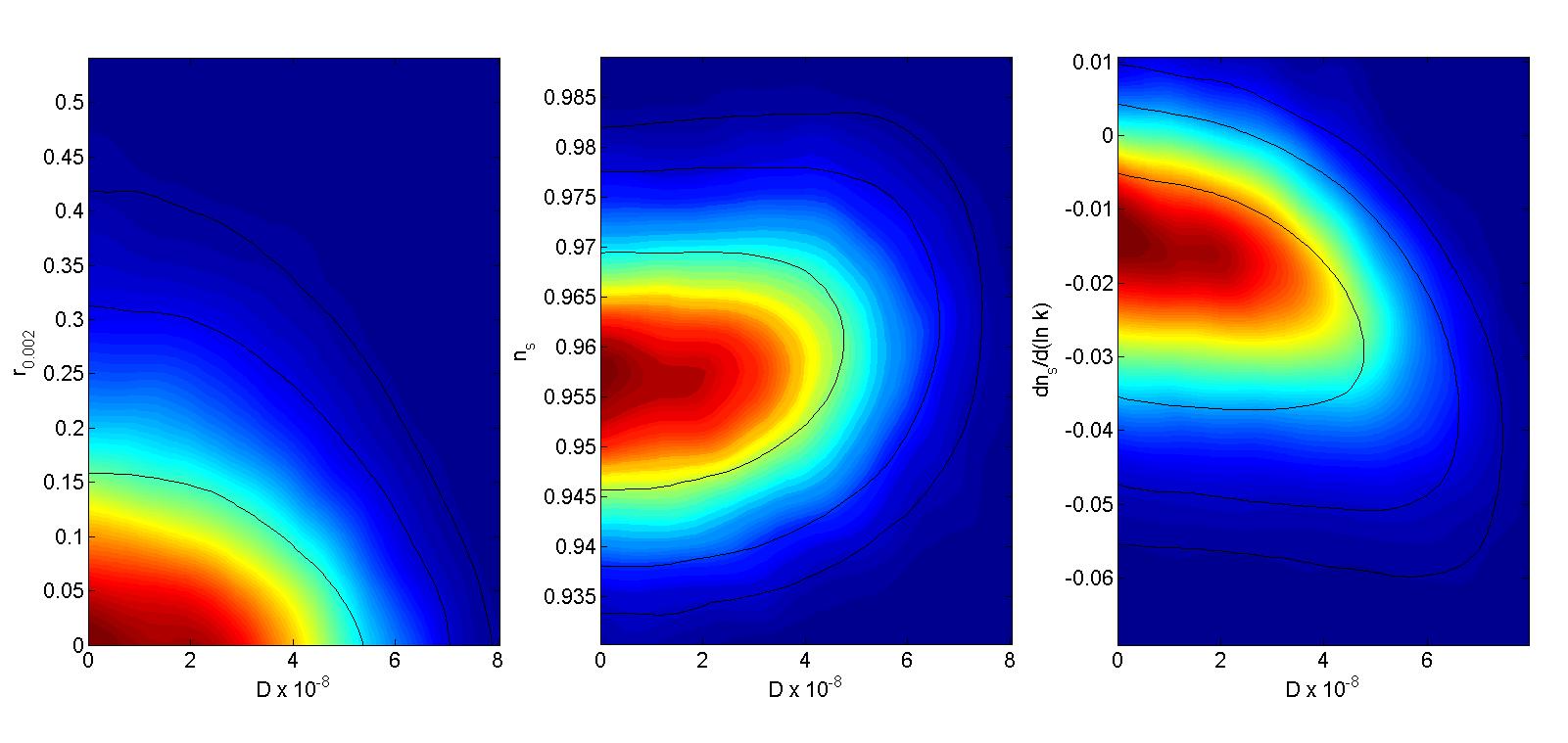}
\caption{Same as in figure~\ref{figu5} for three pairs. Each of them involves $D\times 10^{-8}$ and other 
parameter displayed in the vertical axis of the corresponding panel. 
} 
\label{figu6}%
\end{figure*}

\begin{figure*}                   
\centering
\includegraphics[width=16.cm]{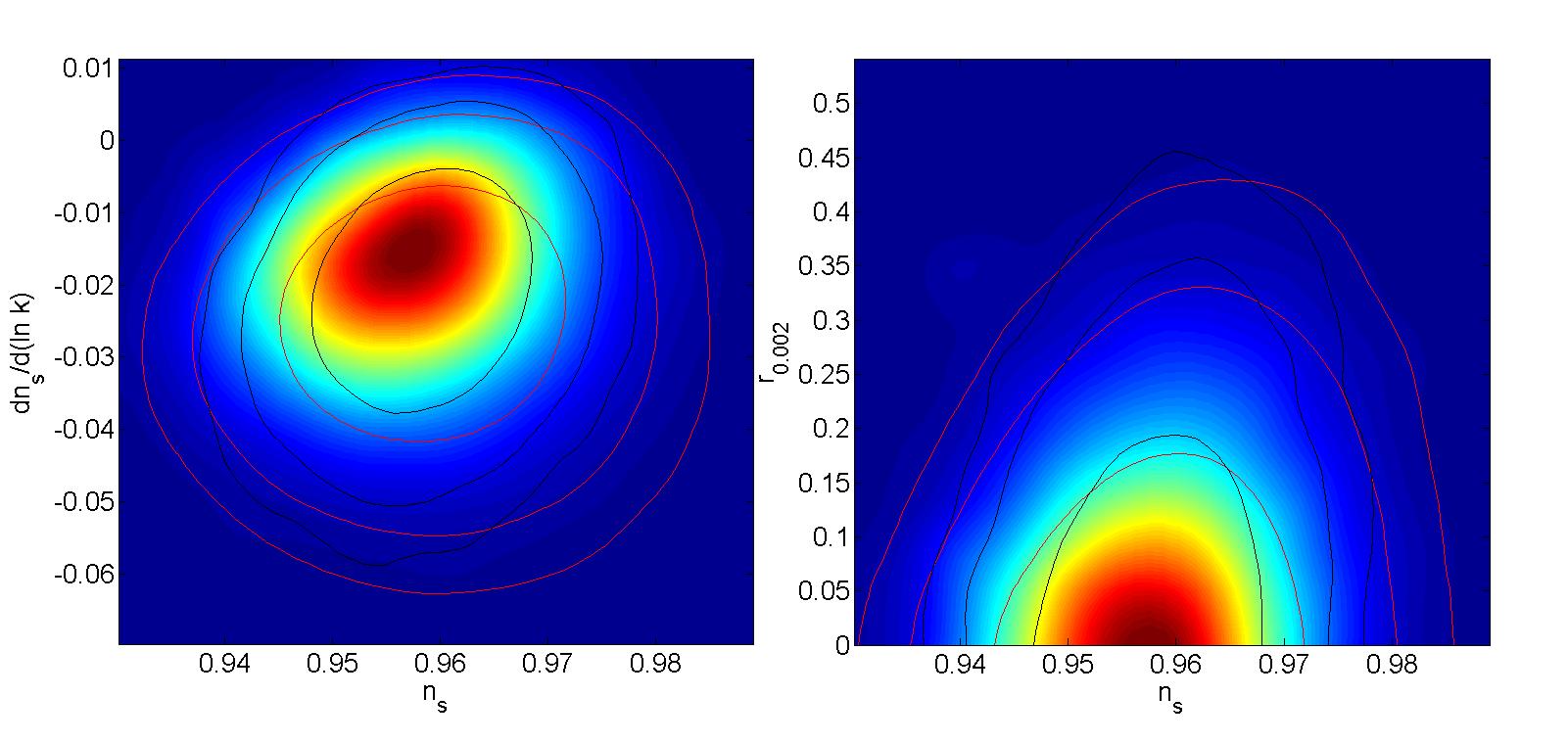}
\caption{Marginalized distribution functions (color) in the Planck+WP-TM-RSI fit 
for the pairs specified in the panels, in which 
red [black] contours correspond to the $1\sigma$ (inner),
$2\sigma$ (middle) and $3\sigma$ (outer) confidence levels in VT [GR]
} 
\label{figu7}%
\end{figure*}

In the context of GR-CM, it has been proved that a running spectral index strongly 
modifies the $r_{0.002} $ upper bound \citep{pla14a} and, on account of this fact, we have studied the Planck+WP-TM-RSI fit.
GR and VT results may be compared by using Table~\ref{table:5}, which has the same structure as 
Table~\ref{table:4}, and figures~\ref{figu3}-\ref{figu4} and~\ref{figu6}-\ref{figu7}.

As it is seen in Table~\ref{table:5}, for the Planck+WP-TM-RSI fit, the $2\sigma $ VT upper limit 
of $D\times 10^{-8}$ is 5.442 and the best fit of this parameter is 1.116; hence, 
the existence of a nonvanising parameter $dn_{s}/d(\ln{k})$ has led to values
of both the best fit and the $2\sigma $ upper limit
greater than those of the Planck+WP-TM fit. A greater upper limit (BF value) is consistent with the fact that 
the dot-dashed line in the right panel of figure~\ref{figu3}
is rather wider (has a wider flat central part) than the dashed one.

Let us now consider parameter 
$r_{0.002}$. In the last row of Table~\ref{table:5},
the relative deviation corresponding to this paramter is $\Delta (U2) \simeq -7.22\%$
and the $2\sigma $ upper limit in VT is $U2 \simeq 0.2574$. The corresponding GR limit 
is then $U2 \simeq 0.2767$ and, consequently, for the Planck+WP-TM-RSI fit, 
one has: 

$r_{0.002} < 0.2574$ (VT, $\simeq 95\%$)

and 

$r_{0.002} < 0.2767$ (GR, $\simeq 95\%$).

This last relation must be compared with eq.~(63b) in \cite{pla14a}, which has the form 
$r_{0.002} < 0.26$ ($\simeq 95\%$) for a Planck+WP+highL-TM-RSI fit in GR. 
The small difference between the values $0.26$ and 0.2767 may be due to  
the use of hihgL data, which are considered only in \cite{pla14a}, and also to the convergence criterium 
which is more severe in our case; in any way, these two values and 0.2574 (VT fit) 
are too similar to speak about significant differences between GR and VT. 

Finally, for the parameter $dn_{s}/d(\ln{k})$, we have found 
$\Delta(L2) = 8.55\%$ and $\Delta(U2) = 51.67\%$ (see Table~\ref{table:5}). 
These are the maximum relative deviations between GR and VT 
arising in this paper.
In the same Table, we also see that, for the parameter under consideration and VT, one has 
$L2 = -0.0482 $ and $U2 = -0.0021$. From all these data one easily finds that the 
corresponding GR values are $L2 = -0.0442 $ and $U2 = -0.0012 $ and, then, 
one can write:

$dn_{s}/d(\ln{k}) = -0.025 \pm 0.023$ (VT, $\simeq 95\%$)    

and

$dn_{s}/d(\ln{k}) = -0.023 \pm 0.021$ (GR, $\simeq 95\%$).

In the same way, at $1\sigma $ confidence, our fit leads to the following 
relations: 

$dn_{s}/d(\ln{k}) = -0.024 \pm 0.012$ (VT, $\simeq 68\%$)

and

$dn_{s}/d(\ln{k}) = -0.021 \pm 0.011$ (GR, $\simeq 68\%$), and this
relation is identical to eq.~(62a) in \cite{pla14a}, which is not 
surprising at all.  

\begin{table*}
\caption{Planck+WP-TM-RSI fit}             
\label{table:5}      
\centering          
\begin{tabular}{l l l l l l l }     
\hline\hline       
Parameter & BF & L2 & U2 & $\Delta(L2)$ & $\Delta(U2)$ & R2\\
\hline                    
$D \times 10^{-8}$ & 1.116 & 0.000 & 5.442 & - & - & - \\  
$\Omega_{b}h^{2}$ & 0.02224 & 0.02175 & 0.02304 & -0.28\% & 0.52\% & 1.162 \\
$\Omega_{c}h^{2}$ & 0.1198 & 0.1139 & 0.1250 & -1.74\% & 1.78\% & 1.609\\
$100\theta_{MC}$ & 1.0411 & 1.0402 & 1.0427& -0.01\% & 0.02\% & 1.136 \\
$\ln{(10^{10}A_{s})}$ & 3.112 & 3.054 & 3.182 & -0.07\% & 0.06\% & 1.032\\
$n_{s} $ & 0.9611 & 0.9426 & 0.9757 & -0.36\% & 0.48\% & 1.324 \\
$\tau $ & 0.0992 & 0.0700 & 0.1344 & -3.23\% & 1.05\% & 1.061 \\
$r_{0.05}$ & 0.0125 & 0.0000& 0.02227& - &-7.39\% & 0.929 \\
$dn_{s}/d(\ln{k})$ & -0.0098 & -0.0482& -0.0021 & 8.55\% & 51.67\%& 1.072\\
\hline
$\Omega_{\Lambda} $ & 0.692 & 0.660 & 0.728 & -2.13\% & 1.68\% & 1.643 \\
$\Omega_{m} $ & 0.308 & 0.272 & 0.340& -4.35\% & 4.27\% & 1.643 \\
$\sigma_{8} $ & 0.849 & 0.818 & 0.874 & -0.20\% & 0.16\% & 1.057 \\
$z_{re} $ & 11.82 & 9.50 & 14.46 & 0.44\% & 0.54\% & 1.007 \\
$H_{0} $ & 67.90 & 65.76 & 70.88 & -1.37\% & 1.46\% & 1.610 \\
$Y_{P} $ & 0.2449 & 0.2447 & 0.2453 & 0.00\% & 0.04\% & 1.200 \\
$t_{0} $ & 13.78 & 13.64 & 13.85 & -0.22\% & 0.14\% & 1.312 \\
$r_{0.002}$ &0.01165 &0.00000 &0.2574 &- &-7.22 & 0.930\\
\hline                  
\end{tabular}
\end{table*}

We have focused our attention on parameters $r_{0.002}$ and  
$dn_{s}/d(\ln{k})$, which are characteristic of the non-minimal fit 
of this section. Hereafter, parameter $n_{s}$ is also considered as it was done in 
section~\ref{sec:4-4}. The marginalized distribution functions $P/P_{max}$ 
of these three parameters
are displayed in figure~\ref{figu4}. 
In the left, central, and right panels, 
the dotted and dashed lines give $P/P_{max}$ in the context of
GR-CM and VT-CM, respectively. By comparing these two types of lines,
one easily concludes that they are rather similar; which means that
the introduction of a running spectral index produces similar effects 
in VT and GR. In both cases, these effects are important as it follows 
from the comparison of dotted with solid lines (GR) and dashed with 
dot-dashed curves in the left and central panels, but this importance is rather 
similar in both theories. All this is in agreement with previous comments
and inequalities based on Table~\ref{table:5}. 

As in section~\ref{sec:4-4}, let us now show the marginalized distribution 
function for the same pairs as in figure~\ref{figu5}, and also for the new pair 
[$D\times 10^{-8},dn_{s}/d(\ln{k})$]. The three functions are displayed in figure~\ref{figu6}.
In the three panels one sees that, if the running spectral index is fitted,
the inequality $|D|\times 10^{-8} \lesssim 5$ is satisfied at the $1\sigma$ 
level, with greater upper limits at $2\sigma$ and $3\sigma$. 
The $|D|$ upper limits of this section are larger than 
those of section~\ref{sec:4-4} where $dn_{s}/d(\ln{k}) = 0$, which means that,
if the running spectral index is fitted, $D$ values larger than those of the 
Planck+WP-TM fit are possible at a given confidence level.

Finally, the marginalized distribution functions of the pairs [$n_{s}$, $dn_{s}/d(\ln{k})$]  
and [$n_{s}$, $r_{0.002}$] are shown in figure~\ref{figu7}, where one easily see 
that the red contours (VT) delimit more extended areas than the black curves (GR) 
for the same confidence level, which suggests once more that a non vanishing 
$D$ parameter facilitates adjustments between theoretical predictions and observational data.

\section{Discussion and conclusions}
\label{sec:5}

Detailed analysis of VT have been developed here and also in 
\cite{dal09,dal12,dal14,dal15}; so, VT has become one of the 
best tested gravity theories. Beside the outcomes described in 
section~\ref{sec:1}, in this paper, we have proved that 
VT explains current CMB anisotropy data due to Planck 
collaboration \citep{pla14a}, and other cosmological observations about 
BAO, SNIa, and so on. 

There are parameters as $\varepsilon$ and $\gamma$ 
involved in the action \ref{1.1}, which keep 
almost arbitrary after our exhaustive analysis.
Only the inequalities $\gamma > 0$ and 
$2 \varepsilon > \gamma$ must be satisfied. The first 
relation is necessary to have positive dark energy 
in the cosmological background (with $W=-1$), and
the second inequality is necessary to built up a VT theory 
without classical instabilities and quantum ghosts. 

The sign of quantity $\Xi_{B}$  -see section~\ref{sec:2}- remains 
arbitrary; however, its absolute value is
fixed to have an admissible $A^{\mu} $ background energy 
density.

Finally, there is another parameter, $D$, 
whose absolute value $|D|$ 
normalizes the 
spectrum of the 
scalar $A^{\mu} $ cosmological perturbations. The 
$|D|$ value control the departures between VT and GR for
scalar cosmological perturbations.
The sign of $D$ is irrelevant.

Cosmological parameters, including $|D|$, have been numerically 
estimated for five significant fits involving different observational data. 
Three of these fits are minimal (seven parameters), whereas other two 
consider additional parameters to study both inflationary gravitational 
waves and running spectral indexes. 
The numerical codes we have 
used -in VT- are suitable modifications of the well known codes CAMB and COSMOMC. 
Our results are similar to those obtained by the Planck collaboration 
in the context of GR \citep{pla14a}, but it has been verified that
parameter $|D|$ 
does not harm the estimation of other parameters involved in the 
standard cosmological model (GR-CM);
on the contrary, if $D$ is considered as an additional parameter
to be adjusted (VT-CM) and a certain confidence level is assumed, 
we have found that, in VT-CM, most
GR parameters belong to intervals wider than those of 
the GR-CM and, consistently, quantity $|D|$ takes on non vanishing 
values. Parameter $|D|$ plays a positive statistical role in
the study of VT scalar perturbations.

New applications or appropriate generalizations of VT could be necessary 
to fix $\gamma $, $\varepsilon $ and the sign of $\Xi_{B}$. 
The new applications should probably be nonlinear as, e.g., the
study a binary stellar systems radiating 
gravitational waves or a deep analysis of the black holes and their
surroundings (see \cite{dal15}). Interesting VT generalizations
could be obtained by replacing $R$ by an
appropriated function $f(R) $ in action~\ref{1.1}; thus, 
the field $A^{\mu} $ could explain the accelerated late time expansion, whereas 
the scalar field, associated to $f(R)$ in the Einstein frame, could account for the 
required inflation; hence, function $f(R) $ would be chosen to achieve a good 
inflation, without producing late time acceleration, which implies less restrictions to 
be satisfied by $f(R)$. Finally,
in appropiate VT generalizations, $A^{\mu}$ vector modes might suitably evolve 
-coupled to other modes of the same type- to explain interesting effects as, e.g., 
the CMB anomalies observed by WMAP and Planck at very large angular scales 
\citep{deo04,cop04,eri04,han04,pla14b}.
This explanation is not easy in the context of GR and VT, where 
vector modes decay \citep{mor07,mor08}.
These promising developments are beyond the scope 
of this paper.

\acknowledgments

This work has been supported by the Spanish
Ministry of Econom\'{\i}a y Competitividad, MICINN-FEDER project
FIS2015-64552-P and CONSOLIDER-INGENIO project CSD2010-0064. 
We thank J.A. Morales-LLadosa
for useful discussion. Calculations were carried out at the 
Centre de c\`alcul de la Universitat de Val\`encia.



\end{document}